\documentstyle[psfig,pstricks,twocolumn,eqsecnum,aps]{revtex}

\begin{document}
\draft
\wideabs{
\title{
Temperature dependent surface relaxations of Ag(111)
}
\author{Jianjun Xie,$^{1}$ Stefano de Gironcoli,$^{2}$
Stefano Baroni,$^{2,3}$  and  Matthias Scheffler$^{1}$}
\address{
$^{1}$ Fritz-Haber-Institut der Max-Planck-Gesellschaft,\\
Faradayweg 4-6, D-14195 Berlin-Dahlem, Germany }
\address{
$^{2}$SISSA -- Scuola Internazionale Superiore di Studi Avanzati and
\\ INFM -- Istituto Nazionale per la Fisica della Materia, \\ via Beirut
2--4, I-34014 Trieste, Italy
}
\address{$^3$CECAM -- Centre Europ\'een de Calcul Atomique et
Mol\'eculaire \\ ENS, Aile LR5, 6, All\'ee d'Italie, 69007 Lyon, France}
\maketitle
\begin{abstract}
The temperature dependent surface relaxation of Ag(111) is calculated
by density-functional theory.  At a given temperature, the equilibrium
geometry is determined by minimizing the Helmholtz free energy within the
quasiharmonic approximation. To this end, phonon dispersions all over
the Brillouin zone are determined from density-functional perturbation
theory. We find that the top-layer relaxation of Ag(111) changes from
an inward contraction ($-$0.8\%) to an outward expansion (+6.3\%) as
the temperature increases from $T=0$  K to 1150 K, in agreement with
experimental findings. Also the calculated surface phonon dispersion
curves at room temperature are in good agreement with helium scattering
measurements.  The mechanism driving this surface expansion is analyzed.
\end{abstract}
}

\section{Introduction}
\label{sec:level1}

The equilibrium geometry of a system depends on the temperature  due
to the anharmonicity of the interatomic potential. The presence of
a surface breaks the periodic structure normal to the surface, and
anharmonic effects are expected to be larger at the surface than in the
bulk\cite{Wette69}. Hence, the surface interlayer separation may change
more strongly with temperature than the bulk lattice parameter. Indeed,
enhanced anharmonic effects have been observed by recent experiments
on several surfaces: Ni(001)~\cite{Cao90}, Pb(110)~\cite{Frenken},
Cu(110)~\cite{Helgesen}, Ag(111)~\cite{Gust94}, Cu(111)~\cite{Gust96} as
well as Be(0001)~\cite{Pohl}. Among them, the large thermal expansion
observed in the close-packed Ag(111) surface has attracted much
attention~\cite{Lewis,Nara,Kara97}, but at present the interpretation
of these results is still controversial.

Using an the embedded-atom method (EAM) in which the parameters of the
interactomic potential are determined by fitting bulk properties,
Lewis~\cite{Lewis} simulated the thermal behavior of Ag(111) for a
large range of temperatures using molecular dynamics. The results
for the top layer relaxation differ significantly from those reported
by an experimental study~\cite{Gust94}: the top interlayer spacing,
$d_{12}$, remains smaller than the bulk value even at temperatures as
high as 1110 K; while the analysis of experimental results~\cite{Gust94}
obtained by the medium energy ion scattering (MEIS), concluded  that
$d_{12}$ changes from $-2.5\%$ contraction to 10.0\% expansion as
temperature increases from room temperature to 1150 K. Narasimhan
and Scheffler~\cite{Nara} investigated the temperature dependence of
$d_{12}$ by minimizing the Helmholtz free energy of the system with
respect to $d_{12}$ in a simplified quasiharmonic approximation (QHA),
where the vibrational free energy was calculated including only three
representative modes corresponding to the rigid vibration of the top
layer on a rigid substrate. The static total energy and the vibrational
frequencies, were calculated using density-functional theory (DFT)
within the local-density approximation (LDA). Although the results
obtained within this ``three-mode approximation'' overestimated the
effect (e.g., at $T=1040$ {\rm K}, the calculated surface relaxation is
$15\%$ whereas the experimental value was 7.5\% ), these calculations
provided a {\em physical explanation} of the mechanism underlying the
thermal expansion observed at this surface.  Subsequently, using again
EAM potential, Kara {\it et al.}~\cite{Kara97} obtained a rather small
thermal expansion. They argued that the large thermal expansion of
Ref.\ \cite{Nara} was the result of an improper representation of
the vibrational density of states. On the other hand, very recent
MEIS measurements on Cu(111)~\cite{Gust96} and LEED measurements on
Be(0001)~\cite{Pohl} seem to support the theoretical picture developed
in Ref.\ \cite{Nara}.

Recent calculations of the thermal properties of Ag bulk~\cite{Xie2}
demonstrate that the QHA provides a very accurate description of the
thermal expansion and heat capacity of Ag up to the melting point.
In order to resolve the controversy on the thermal behavior of Ag(111),
we have recalculated the surface thermal expansion of this surface
within DFT-LDA and QHA without any further approximations. In particular,
the vibrational contributions to the free energy from the whole BZ are
included thanks to the efficient calculation of phonon dispersions by
density-functional perturbation theory~\cite{Baroni87}.  Our results
positively indicate that DFT and the QHA---at variance with previous
attempts based on EAM~\cite{Lewis,Kara97}---provide a {\it quantitatively}
correct description of the anomalous thermal properties of this surface.
Our results also 
show the importance of a proper sampling of vibrational modes over the
BZ for a quantitatively reliable result. The {\em qualitative} explanation
of the earlier work of Narasimhan and Scheffler \cite{Nara}
is fully confirmed. The disagreement with reported EAM results
\cite{Lewis,Kara97}  is argued to due to the approximate nature of EAM 
and to an incorrect {\bf k}-point summation in Ref. \cite{Kara97}.

\section{Computational details} \label{sec:level2} 
To model the surface, we adopt a repeated-slab geometry consisting
of seven atomic layers separated by a vacuum region corresponding
to five atomic layers. As in a previous treatment~\cite{Xie1}, the
Helmholtz free energy of the slab is given by 
\\ \begin{eqnarray}
F(\{d\},T)&=&E(\{d\})+F_{\rm vib}(\{d\},T) \nonumber\\
&=&E(\{d\})+  \nonumber\\
k_{\rm B}T\sum_{{\bf q}_\parallel}\sum_{p=1}^{3N}&& {\rm
ln}\left\{2\sinh\left(\frac{\hbar\omega_{p}({\bf q}_\parallel,\{d\})}
{2k_{\rm B}T}\right)\right\} \quad, 
\end{eqnarray} 
where $E$ is the static
total energy, as obtained by DFT calculations, $k_{\rm B}$ and $\hbar$
are the Boltzmann and the Planck constants, and \{$d$\} represents the set
of interlayer distances normal to the surface and  the inter-particle
distances parallel to the surface.  The vibrational free energy is
denoted as $F_{\rm vib}$, and $\omega_{p}({\bf q}_\parallel,\{d\})$
is the frequency of the $p$-th mode for a given wave vector {\bf
q}$_\parallel$, evaluated at the geometry defined by \{$d$\}; and $N$
is the number of atoms in the slab.  The static total energy  $E$ in Eq.\
(2.1) includes  all the anharmonic terms  of the interatomic potential.
The anharmonic nature also appears in the vibrational free energy $F_{\rm
vib}$ since in the quasiharmonic approximation the vibrational frequencies
$\omega_{p}({\bf q}_\parallel,\{d\})$ are allowed to change with $\{d\}$.

At a given temperature $T$ and zero pressure, the equilibrium geometry
is determined by the minimum of the
Helmholtz free energy, i.e., $\partial
F/\partial d$ = $\partial E/\partial d$ +$\partial F_{vib}/\partial
d=0$. It is not practical to minime the Helmholtz free energy with
respect to all the lattice parameters $\{d\}$ within the present {\it
ab initio} approach.  We therefore assume here that the vibrational free
energy $F_{vib}$ depends parametrically only on the interlayer distance
between the top layers, $d_{12}$. All other interlayer distances and
inter-particle distances are assumed to vary with temperature as in the
bulk~\cite{Xie2}.  The equation which determines the temperature dependent
equilibrium $d_{12}$, then reads\ \begin{eqnarray} \frac{\partial
E(d_{12})}{\partial d_{12}}=-\frac{\partial F_{\rm vib}} {\partial
d_{12}}=- \sum_{{\bf q}}\sum_{p} \frac{\partial \hbar \omega_{p}({\bf q})}
{\partial d_{12}}n_{p}({\bf q}) \quad , \end{eqnarray} where  $n_{p}$
is the occupation number of the $p$-th mode defined by \begin{equation}
n_{p}=\frac{1}{2}+ \frac{1}{e^{\hbar\omega_{p}({\bf q})/k_{\bf B}T}-1}.
\end{equation}

The static total energy $E$ is calculated by density-functional theory
within the local-density approximation~\cite{Ceperley}.  Fully separable
norm-conserving pseudopotentials~\cite{Gonze} are used in our calculations
together with a plane wave basis set with a kinetic energy cut-off of
55 Ry. BZ integrations are performed with the smearing technique of
Ref.~\cite{Methfessel} using the Hermite-Gauss smearing function of
order one, a smearing width $\sigma=50$ mRy, and a 16-point grid in the
irreducible wedge of the BZ. The phonon frequencies of the system are
calculated by density-functional perturbation theory~\cite{Baroni87}.
The dynamical matrices are calculated on a 4$\times$4 grid of points
in the surface BZ of the 7-layer slab and Fourier interpolated over a
48-point grid of ${\bf q}_{||}$ vectors in the irreducible wedge of the
surface BZ, in order to calculate the vibrational contribution to the
free energy.

\section{Results}
We first calculated the surface relaxation of Ag(111) by minimizing
the static total energy and neglecting the vibrational contributions.
The obtained results are : $\Delta d_{12}/d_{\rm B}=-1.0\% $, $\Delta
d_{23}/d_{\rm B}=-0.2\%$ , where $d_{\rm B}$ is the interlayer spacing in
the bulk. Starting from this static equilibrium geometry, the Helmholtz
free energy of the slab is evaluated as a function of the top layer
interspacing $d_{12}$.  The temperature dependence of the equilibrium
$d_{12}$ is then obtained from Eq.\ (2.2).  All other interlayer
spacings are assumed to expand according to the temperature dependence
of the bulk.  
\begin{figure}[tbh]
\pspicture(0,0)(8,7)
\rput[l](-1.5,2.5){\psfig{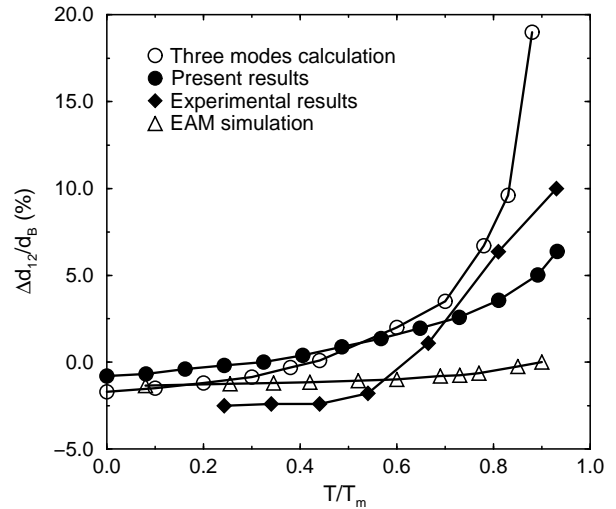}}
\endpspicture
\caption{Temperature dependence of surface layer relaxation of Ag(111).
Our calculated results are denoted by the filled circles. Open circles 
are the results of Ref.~{\protect \cite{Nara}}, open triangles are EAM 
simulations~{\protect \cite{Kara97}}. The experimental 
results~{\protect \cite{Gust94}} are shown by filled diamonds.}
\label{figure1}
\end{figure}
Figure 1 shows the calculated results  (filled circles)
together with the experimental data~\cite{Gust94} (filled diamonds)
and other calculations~\cite{Nara,Kara97}. The temperature is scaled
with respect to the experimental melting temperature ($T_{\rm m} =
1234$ K\cite{Kittel} ).  It can be seen that in the low temperature
range ($T/T_{\rm m} < 0.6 $), the present calculation gives a thermal
expansion similar to that of Ref.~\cite{Nara}. At higher temperatures,
the calculations of Ref.~\cite{Nara} overestimate the thermal expansion
of Ag(111).  The present calculation, which includes all the vibrational
modes of the slab in the whole BZ, displays a much smaller top layer
expansion than obtained in the ``three-mode approximation" \cite{Nara},
thus bringing the theoretical predictions in much closer agreement with
experimental results: $\Delta d_{12}/d_{\rm B}$ changes from $-$0.8\%
inward contraction (including zero-point vibrations) at $T=0$ K to +6.3\%
outward expansion at $T=1150$ K (the corresponding experimental figures
vary from $-$2.5\% to $+10.0$\%~\cite{Gust94}).  The EAM simulation in
Ref.~\cite{Kara97}, on the contrary, shows no enhancement of thermal
expansion of the interlayer spacing in the whole temperature range,
{\em i.e.}, the $d_{12}$ value  always remains smaller than the 
interlayer spacing in the bulk.

\begin{figure}[tbh]
\pspicture(0,0)(8,8)
\rput(4,2.5){\psfig{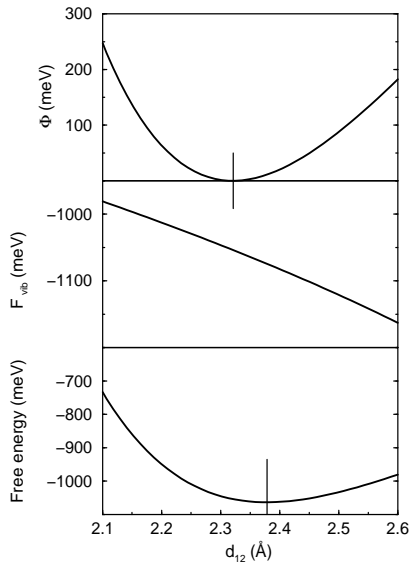}}
\endpspicture
\caption{Variation of the static total energy $E$, vibrational free energy
$F_{\rm vib}$ and the Helmholtz free energy with the surface interlayer spacing
$d_{12}$ at $T$=500  K.
}
\label{figure2}
\end{figure}
What is the mechanism giving Ag(111) such an enhanced thermal expansion?
Figure 2 shows the variation of the static energy, $E$, the Helmholtz
free energy, $F$, and the vibrational free energy, $F_{\rm vib}$, with
the top-layer interspacing $d_{12}$, for $T = 500$ K. The equilibrium
geometry, $d_{12}$, is determined by two factors: one is the static total
energy, $E$, which governs the binding strength of the surface with the
substrate and the anharmonicity of the interlayer potential normal to
the surface; the other is  the decrease of the vibrational free energy,
$F_{\rm vib}$, which reflects the ``softening'' of the vibrational
frequency with the increase of $d_{12}$. We note that $F_{\rm vib}$
does not always decrease with the increase of lattice parameters, for
example, in the anomalous thermal expansion of bulk 
silicon~\cite{Biernacki89}.
As $d_{12}$ increases  from the static equilibrium value, $d_{12}^{0}$,
the interlayer potential increases while the vibrational free energy
decreases, determining a new equilibrium spacing $d_{12}$, larger than
$d_{12}^{0}$.  
\begin{figure}[tbh]
\pspicture(0,0)(8,8)
\rput(4.5,2.7){\psfig{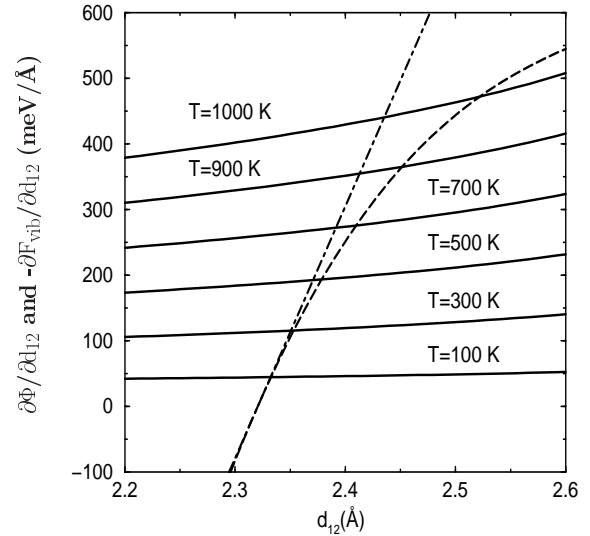}}
\endpspicture
\caption{Variation of $-\partial F_{\rm vib}/\partial d_{12}$ 
(solid lines) and $\partial E/\partial d_{12}$ (dashed line)
 with $d_{12}$. The harmonic results of $\partial E/\partial d_{12}$
are shown by the dot-dashed line. The equilibrium $d_{12}$ is obtained
by the intersection of the solid lines and the dashed line.
}
\label{figure3}
\end{figure}  
Figure 3 shows how the temperature dependent equilibrium
$d_{12}$ is obtained from Eq. (2.2). The dashed line is the variation of
$\partial E/\partial d_{12}$ with  $d_{12}$. The dot-dashed line is the
derivative of $E$ with respect to $d_{12}$ when anharmonic terms in $E$
are neglected by quadratically expanding it around the static equilibrium
geometry.  The solid lines are $-\partial F_{\rm vib}/ \partial d_{12}$
at different temperatures. The equilibrium geometry of $d_{12}$ is
determined by the intersection of  $\partial E/\partial d_{12}$ and
$-\partial F_{\rm vib}/\partial d_{12}$. It can be clearly seen that
by increasing the temperature, the intersection of $\partial E/\partial
d_{12}$ and $-\partial F_{\rm vib}/\partial d_{12}$ gives an increasing
value of equilibrium distance $d_{12}$.  Furthermore, the anharmonicity
of the static energy $E$, dashed line, is essential in determining
the enhanced thermal expansion at high temperature, that would be much
reduced if this anharmonicity were neglected (dot-dashed line).

The driving force for expansion comes from the temperature variation
of the vibrational free energy. Equation (2.2) reveals that the value
of $-\partial F_{\rm vib}/\partial d_{12}$ at a given temperature
is determined by $\partial\hbar\omega_{p}/\partial d_{12}$ which
represents the ``softening'' of $\omega_{p}$ with the increase of
$d_{12}$.  Such ``softening" of the frequencies is characterized by
the shift of the phonon density of states (DOS).
\begin{figure}[tbh]
\pspicture(0,0)(8,9)
\rput(4.5,4){\psfig{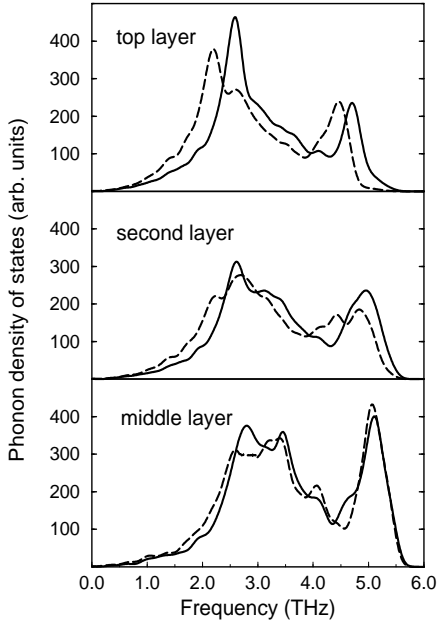}}
\endpspicture
\caption{The phonon density of states (DOS) of different atomic layers.  
Solid lines are for $d_{12}$=2.32 \AA, dashed lines are for $d_{12}$=2.42
\AA.}
\label{figure4}
\end{figure}  
Figure 4 shows the
shift of layer localized  phonon DOS with $d_{12}$. The solid line
corresponds to $d_{12}=2.32$ \AA\ and the dashed line corresponds
to $d_{12}=2.42$ \AA.  Clearly, the DOS of the middle layer of the
slab is quite bulk like, and the shift with $d_{12}$ is very small,
while the surface-layer DOS is significantly different from that of
the middle layer and is very sensitive to $d_{12}$.  By increasing
$d_{12}$, the vibrational frequencies of the first layer ``shift"
downward, which results in a decrease of the vibrational free energy
as shown in Fig.\ 2. The vibrations of atoms in the second layer also
``soften" by increasing $d_{12}$ and contribute to the decrease of the
vibrational free energy. 
\begin{figure}[tbh]
\pspicture(0,0)(8,8)
\rput(4.5,3){\psfig{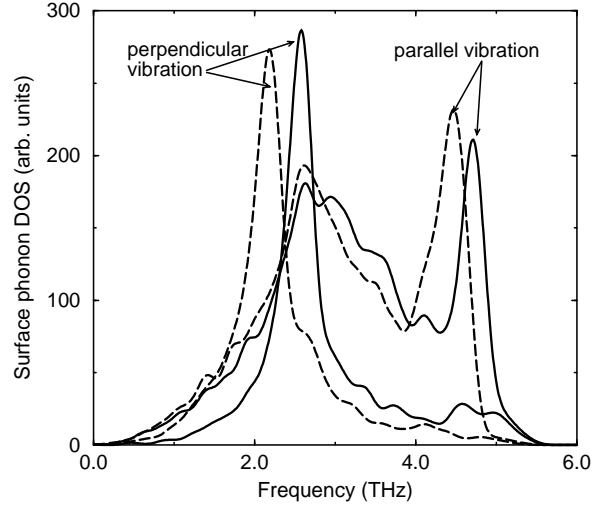}}
\endpspicture
\caption{The shift of phonon DOS at surface from different vibrational modes
with different $d_{12}$. The solids are for $d_{12}$=2.32 \AA and dashed lines
 are for $d_{12}$=2.42 \AA. }
\label{figure5}
\end{figure} 
Figure 5  shows the ``shift" of the phonon DOS
in the top layer with $d_{12}$ for different vibrational modes.  It can
be seen that as $d_{12}$ increases from 2.32 \AA\, to 2.42 \AA, not only
the vibrational modes that are perpendicular to surface, but also those
parallel to surface, shift downward. The shift of the perpendicular modes
reflects the anharmonicity of the interlayer potential $E$ normal to the
surface, while the shift of the parallel modes is due to the flattening
of the corrugation of the potential parallel to the surface. The latter
effect implies that anharmonicity of the interatomic potential at a
surface is a somewhat complicated effect where the $x$, $y$, and $z$
degrees of freedom are not independent.  The modes perpendicular to the
surface (corresponding to the Rayleigh wave) are mainly located in the
low frequency range and provide a smaller contribution to the DOS. The
parallel modes have relatively higher frequencies and occupy a larger
portion of the total DOS.

As discussed above in Fig. 2, the increase of $-\partial F_{\rm
vib}/\partial d_{12}$ with temperature determines the expansion of
$d_{12}$. 
\begin{figure}[tbh]
\pspicture(0,0)(8,8)
\rput(4.5,2){\psfig{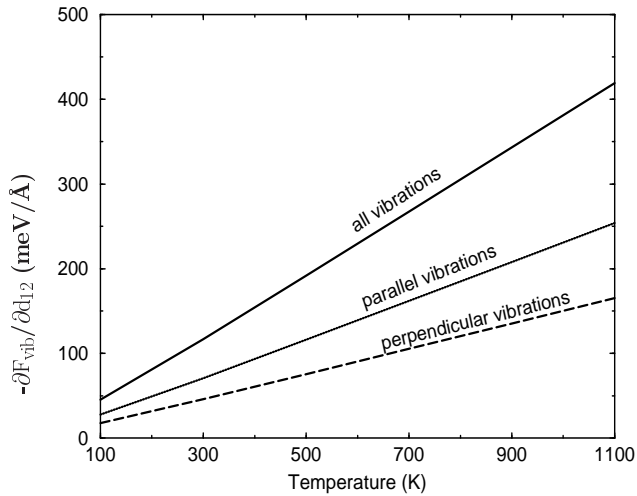}}
\endpspicture
\caption{Contributions to the vibrational free energy from different
vibrational modes}
\label{figure6}
\end{figure}
In Figure 6 the different contributions to $-\partial F_{\rm
vib}/\partial d_{12}$ coming from perpendicular and parallel modes are
analysed. It can be seen that the contribution of the parallel modes
to $-\partial F_{\rm vib}/\partial d_{12}$ is larger than the one of
perpendicular modes and that the difference between the contributions
from parallel and perpendicular modes increases with temperature. This
is in agreement with previous findings~\cite{Nara,Xie1}. Note, however,
that this effect is not general and in other systems, as for instance
the Be(0001) surface \cite{Lazzeri}, the ``softening" of parallel modes
plays a minor role.
Altogether, the enhanced thermal expansion of $d_{12}$ at Ag(111) surface
is governed  by the anharmonicity of the interlayer potential normal to
the surface as well as the ``softening'' of the parallel modes with the
increase of $d_{12}$.

\begin{figure}[tbh]
\pspicture(0,0)(8,7)
\rput(4,1.5){\psfig{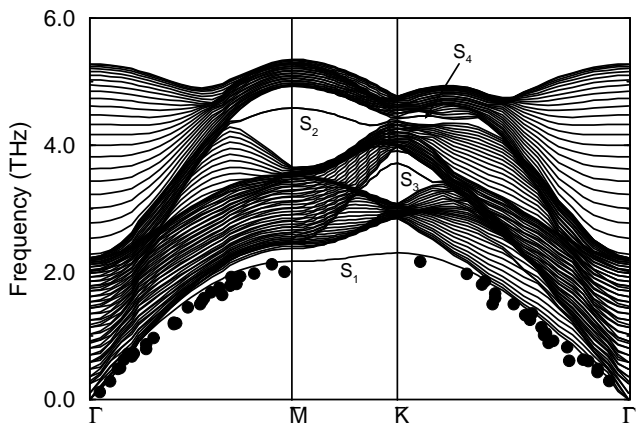}}
\endpspicture
\caption{Calculated phonon dispersion of a 30-layer slab modeling an Ag (111) 
surface.}
\label{figure7}
\end{figure}
Finally we show in Figure 7 the surface phonon band structure
corresponding to the geometry obtained for $T=300$ {\rm K}.  The thickness
of the slab has been extended to 30 atomic layers in order to decouple the
surface vibrations that penetrate deeply in the bulk, by inserting in the
middle of our slab a number of layers with bulk-like force constants. The
lowest surface-mode branch $S_{1}$ which lies below the bulk bands is
the Rayleigh  wave.  It is primarily associated with vibrations normal
to the surface ( shear-vertical (SV) mode, compare also Fig. 5). The
gap mode $S_{3}$ is primarily a ``shear-horizontal" (SH) mode which is
associated with vibrations in the direction transverse to {\bf q$_{||}$}
and parallel to the surface. The surface modes $S_{2}$ and $S_{4}$ are
primarily longitudinal modes which are associated with vibrations along
the direction parallel to {\bf q}$_{||}$.  The calculated frequencies
of the Rayleigh mode $S_{1}$ are in good agreement with  experimental
data from helium scattering\cite{Toennies}.

\section{Conclusion} Our calculation of the thermal properties of Ag(111)
surface are in quantitative agreement with the enhanced thermal expansion
found in the experiments. This behavior is found to be determined by
two effects: $i)$ the anharmonicity of the interlayer potential normal
to the surface and $ii)$ the decrease of the vibrational free energy
with increasing $d_{12}$. The latter effect reflects not only the
anharmonicity of the interlayer potential normal to the surface, but also
the flattening of the corrugation of the interlayer potential parallel
to the surface with the increase of $d_{12}$.
A recent calculation using the 
EAM~\cite{Kara97} shows that the interlayer spacing $d_{12}$ increases
very weakly between $T = 0 $ K and 1100 K. This significant difference
to our {\it ab initio} result is due to two reasons: $i)$ The 
{\bf k}-summation
in Ref.\cite{Kara97} was done incorrectly, i.e., the ratio of weights
at $\bar{\Gamma}$, $\bar{K}$ and $\bar{M}$ should be 1:2:3, 
not 1:6:6, and at the fcc (111) surface  the vibrational modes at
$\bar{\rm M}$ and $\bar{\rm K}$ are different.  $ii)$ the EAM 
lacks some important aspects of the quantum mechanical 
nature of electrons~\cite{Refa}.

\acknowledgments

One of the authors (J.J. Xie) would like to acknowledge the financial
support from Alexander von Humboldt foundation in Germany.  Two of us
(SB and SdG) have done this work in part within the {\sl Iniziativa
Trasversale Calcolo Parallelo} of INFM.  We thank P.\ Ruggerone for
helpful discussions.


\begin{references}
\bibitem{Refa}
``Embedded atom method'' is one name out of several names for essentially
the same  approach. Others are e.g. bond cutting model, effective medium
theory, glue model. All these ``bond cutting models" neglect in their 
standard applications the shell structure of atoms ($s$, $p$, $d$, ...
electrons)
and describe the interaction between atoms by a sum of pair potential
and effective atomic-many-body contributions. Thus, when the kinetic energy
of electrons and/or  a hybridization of $s$, $p$, $d$, ... orbitals
is changing, ``bond cutting models" are likely to fail.
\end{references}
\end{document}